\documentclass[a4paper]{article}

\usepackage{caption}
\usepackage{color}
\usepackage{subcaption}
\usepackage{spconf,amsmath,amsfonts,graphicx}
\usepackage{multirow}

\usepackage{hyperref}

\usepackage{INTERSPEECH2022}

\newcommand{\tuneenv}{SoundFilter}
\newcommand{\soundwords}{SoundWords}

\title{Text-Driven Separation of Arbitrary Sounds}

\name{Kevin Kilgour, Beat Gfeller, Qingqing Huang, Aren Jansen, Scott Wisdom, Marco Tagliasacchi}
\address{Google Research}
\email{\{kkilgour, beatg, qqhuang, arenjansen, scottwisdom, mtagliasacchi\}@google.com}

\ninept

\begin{document}


\maketitle
\begin{abstract}
  We propose a method of separating a desired sound source from a single-channel mixture, based on either a textual description or a short audio sample of the target source. This is achieved by combining two distinct models. The first model, \emph{SoundWords}, is trained to jointly embed both an audio clip and its textual description to the same embedding in a shared representation. The second model, \emph{\tuneenv}, takes a mixed source audio clip as an input and separates it based on a conditioning vector from the shared text-audio representation defined by SoundWords, making the model agnostic to the conditioning modality. Evaluating on multiple datasets, we show that our approach can achieve an SI-SDR of 9.1 dB for mixtures of
two arbitrary sounds when conditioned on text and 10.1 dB when conditioned on audio. We also show that SoundWords is effective at learning co-embeddings and that our multi-modal training approach improves the performance of \tuneenv.
 \end{abstract}

\section{Introduction}

Imagine walking past a pub in small town around dusk. You hear a football commentator's excited chatter suddenly get drowned out by cheering, a car is driving up the road with obnoxious music blaring out of their open windows, while somewhere behind you a cyclist rings their bell; across the road someone is still hard at work in their workshop drilling some holes into a wall, and in the distance you can hear the hooting of an owl. People are very good at picking out the sounds we care about and ignoring the rest. Computers, on the other hand, struggle immensely at this task, especially if the audio is only provided as a single channel input and the goal is to focus on an arbitrary sound, e.g., \textit{the music from the car} or \textit{traffic noise}.

If the type of audio to focus on is known in advance, dedicated enhancement algorithms can be trained to separate out that audio and suppress everything else~\cite{reddy2021interspeech}. In~\cite{weninger2015speech} and~\cite{li2021icassp}, for example, systems are built to enhance speech whereas~\cite{defossez2019music} and~\cite{lluis2018end} demonstrate methods of extracting the audio of individual musical instruments. On the other extreme, some research~\cite{kavalerov2019universal, tzinis2020improving} has been done on the problem of universal sound separation, where the goal is to
split a single channel audio mixture into its individual component signals regardless of their class. These methods could be used and followed by a subsequent selection step that picks which component signal the user wants to focus on, e.g.,~\cite{tzinis2021into}. While this approach is  more versatile than having a set of predefined target classes, it requires to estimate each separate sound in the mixture, making the user interface more complicated and requiring more computational resources compared to enhancing only one type or class of target sound.

In this paper we propose a way of controlling the output of a universal sound separation model by
simply providing a text description as conditioning to tell the model what to focus on.

Some previous approaches have used either images~\cite{gao2019co,xu2019recursive} or audio~\cite{gfeller2021one} as the conditioning modality. 
A downside of both of these is the ambiguity of
how specifically the model should interpret the provided conditioning. For example, if provided with an image of an owl hooting, should the model focus on all bird sounds, just on sounds made by owls, or only on this particular individual? Or maybe it should focus on hooting sounds, but not other sounds such as flapping made by this owl.  The old saying, \textit{a picture is worth a thousand words}, does not apply here and instead a succinct description, e.g., \textit{an owl hooting} is instead worth many pictures.
The same issue is present when conditioning on audio, which has some additional complications, as some types of sound change greatly over time. For example, the starting noise of an engine is not the same as its constant purr when running.

We address this by proposing a model that can be conditioned on arbitrary textual descriptions of sound, while still allowing audio conditioning as an alternative. As we will see in Section~\ref{sec:eval}, if we train a model using only textual description captions, then due to the limited amount of data we run into overfitting issues. To alleviate this problem we begin by training an embedding model that co-embeds audio clips and their textual descriptions close together in same embedding space. Using this embedding model, we can then train our audio separation model using both audio and text conditioning.

\begin{figure*}[ht]
    \centering
    \includegraphics[width=0.9\textwidth]{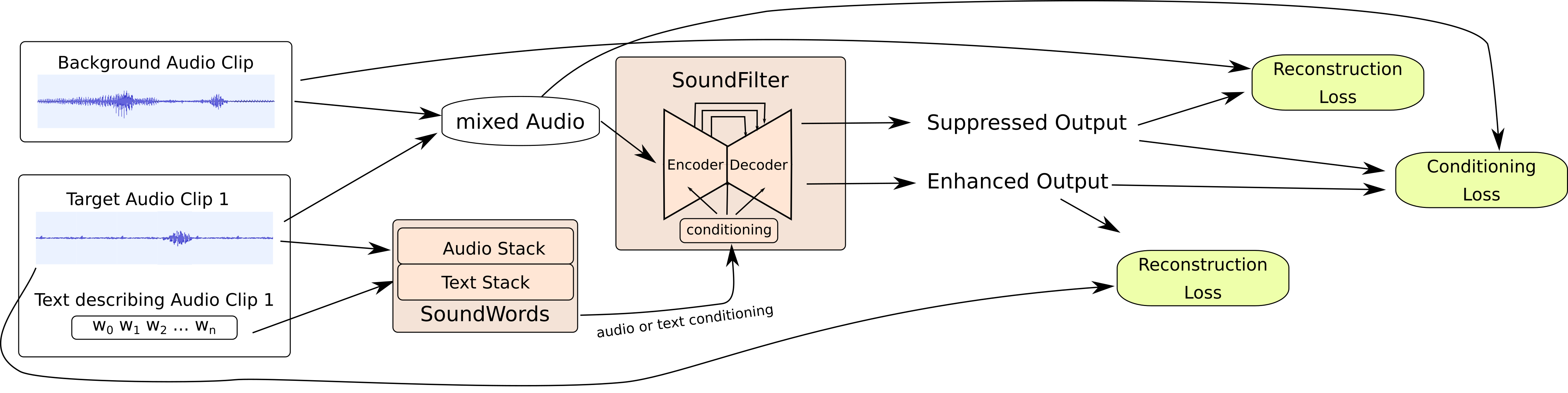}
    \caption{Schematic overview of the training setup.}
     \label{fig:training}  
\end{figure*}


\section{Related Work}
\label{sec:relatedwork}


Universal sound separation, the task of separating all sounds from an acoustic mixture regardless of class, was recently shown to be possible with supervised data \cite{kavalerov2019universal}. Since then, more weakly supervised methods have been proposed, such as using sound class as weak labels \cite{pishdadian2020finding, kongIcassp2020}, and even completely unsupervised methods that can learn directly from raw mixtures of sounds \cite{wisdom2020unsupervised}.

A number of works have suggested methods of conditioning the behavior of neural sound separation models in order to both control the models' behavior and exploit additional information present in modalities besides audio.
In particular, audio-visual methods~\cite{michelsanti2021overview} have been proposed for more specific tasks like speech enhancement~\cite{ephrat2018looking}, and separating limited sets of classes such as musical instruments~\cite{gao2019co, xu2019recursive}. More recent audio-visual work~\cite{tzinis2021into} has provided truly general audio-visual universal separation models that can separate all sounds that originate from visible on-screen objects.
Other examples of conditioning inputs include accelerometer data to improve speech enhancement~\cite{tagliasacchi2020seanet}, target speaker embedding for speech separation~\cite{wang2019voicefilter}, or sound class for universal separation, using either a vector indicating desired source class(es)~\cite{tzinis2020improving, ochiai2020listenToWhatYouWant, zero-shot-audio-source-separation-2022} or an audio clip from a desired class of sound~\cite{gfeller2021one}. Conditioning is generally done by injecting the separation network itself with embeddings extracted from the conditioning input, which is the approach we take in this paper. 

Text conditioning for separation models has received very limited attention. The most related work has been for more restrictive speech applications where the text corresponds to the transcript of a target speaker, such as text-informed nonnegative matrix factorization for speech separation \cite{le2015text} or lyric-informed singing voice separation \cite{schulze-forster2021phoneme}. To the best of our knowledge, we believe our method is the first text-driven neural universal separation model, where text is a natural language description of the sound(s) to extract.

There has been a recent surge of efforts in building joint image and natural
language embedding models in the computer vision community.  Rather than relying
on fixed ontologies, these models provide a natural interface to support
cross-modal retrieval and zero-shot classification applications~\cite{bib:clip,
  bib:align}.  To the best of our knowledge, there has only been one analogous
effort in the sound event community, where a joint embedding model was built
on top of separately-trained and fixed neural audio and text
embeddings~\cite{bib:vgg}.  This joint embedding was evaluated on a cross-modal
retrieval task, but it was not considered in any separation or generative audio
applications.

\section{Method}

Building upon previous work \cite{gfeller2021one}, our model consists of two components: 
(a) a \emph{conditioning encoder}, which takes a conditioning input in the form of either text or audio and computes the corresponding embedding, and
(b) a \emph{conditional generator}, which takes the mixture audio and the conditioning embedding as input, and produces the filtered output.
The two models are trained separately, as (a) requires audio examples with text descriptions, whereas (b) can be trained either with text or with audio as conditioning.


\subsection{Model architecture}

\begin{figure}
    \centering
    \includegraphics[trim={0.125cm 0 0.1cm 0},clip,width=0.495\linewidth]{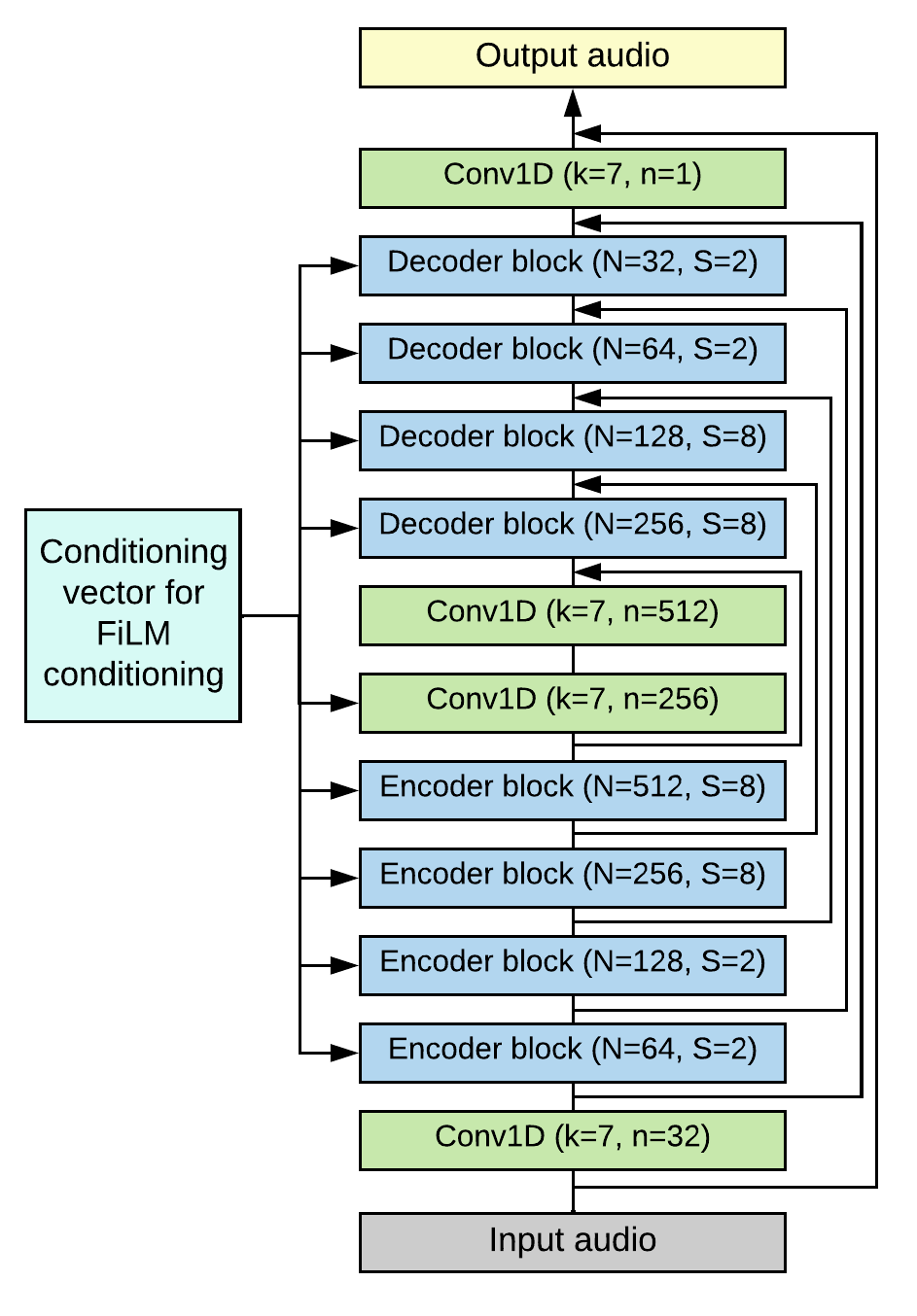}
    \includegraphics[trim={0.125cm 0 0.25cm 0},clip,width=0.495\linewidth]{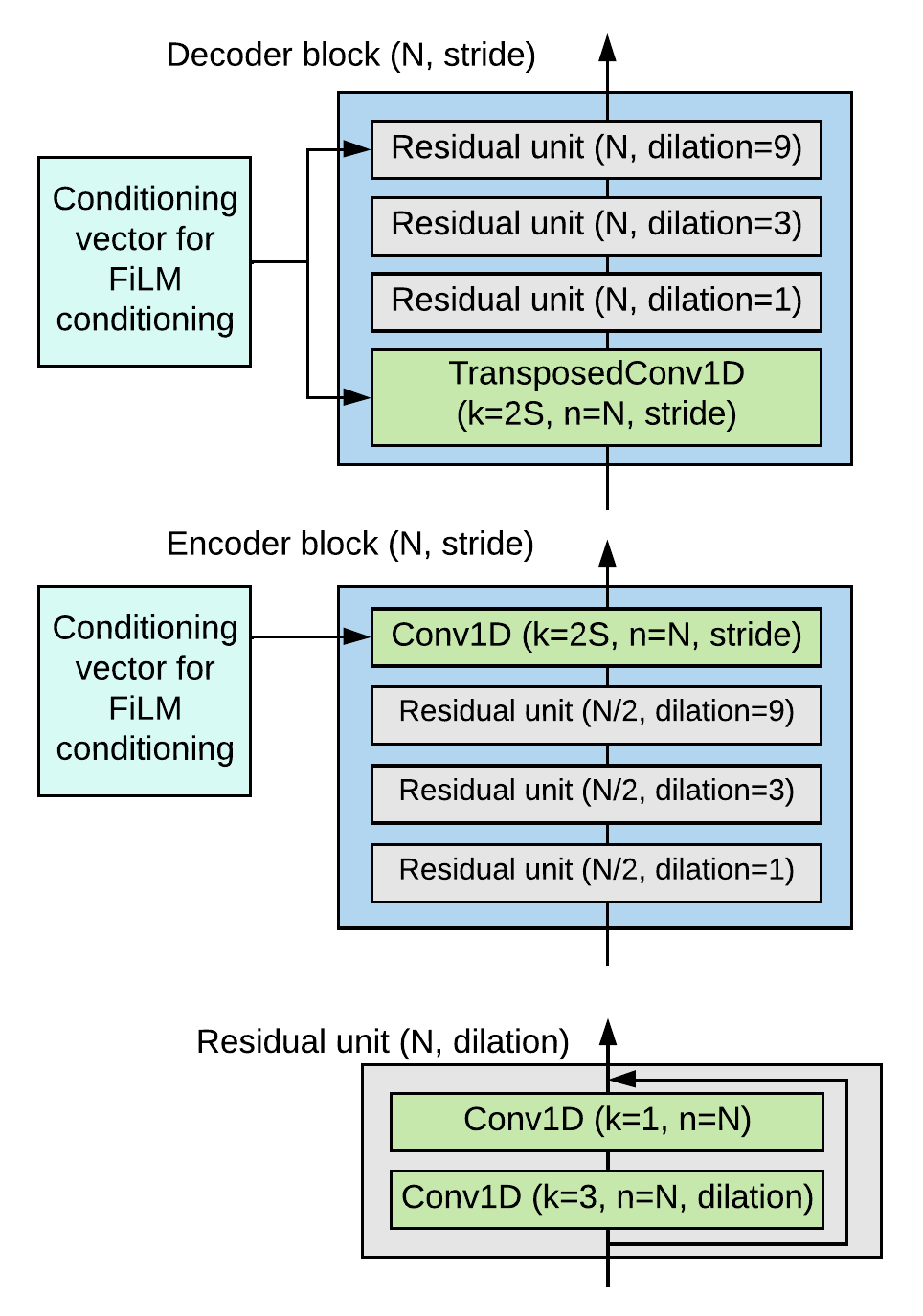} 
    \caption{Detailed outline of the conditioned audio-to-audio U-Net architecture.}
        \label{fig:unet_architecture}
\end{figure}

For (a) we built SoundWords, a joint embedding model of natural language and sound
 that consists of dedicated towers for each, but which are trained to
terminate in the same target space.  In a similar manner to~\cite{bib:align},
this was accomplished through contrastive learning using a large set of audio
clips paired with associated natural language descriptions.  By applying the
contrastive multiview coding loss function~\cite{bib:cmc}, the learned
embedding acts to colocate audio and its underlying semantic categories expressed via
freeform natural language.



For the audio tower, we choose the commonly-used Resnetish-50
architecture~\cite{bib:shershey}, trained on top of 64-channel log mel
spectrograms.  Instead of the usual classifier head, we apply a fully-connected
final layer with the number of units defining the ultimate embedding dimension.  The model
takes as input 10-second clips and applies average pooling on the frame-level
outputs to produce a single clip-level embedding.  For the text tower, we use
the standard BERT-base transformer architecture~\cite{bib:bert}, using
[CLS]-token pooling and a final fully-connected layer to map to the shared
embedding space.

For (b), we use the same architecture as used in previous work \cite{gfeller2021one}:
a symmetric encoder-decoder U-net network with skip-connections, operating in the waveform domain, where the architecture of the decoder layers mirrors the structure of the encoder, and the skip-connections run between each encoder block and its mirrored decoder block (see Figure~\ref{fig:unet_architecture}).
Conditioning is introduced into the model via 
feature-wise linear modulation (FiLM)
conditioning~\cite{perez2017film,DBLP:conf/nips/BirnbaumKEKE19} which is applied to some
layers in the encoder, decoder and bottleneck, as indicated in Figure~\ref{fig:unet_architecture}. We apply group normalization after each Conv1D operation (not shown in the figure for simplicity)~\cite{groupnorm}, and use the ELU activation function~\cite{clevert2016fast} for all non-linearities in the \tuneenv\ model. We refer to \cite{gfeller2021one} for more details on the architecture.

\subsection{Data}

We train \soundwords\ on a combination of several (audio, text)
pair data sources, covering a wide range of scales, supervisory quality, and
language usage patterns.  The first and largest data source is a collection of
approximately 50 million 10-second sound clips extracted from randomly selected
internet videos (1 clip/video). Here, the associated text is the names and
natural language descriptions of the knowledge graph entities associated with
each video.  While many of these entities are directly or indirectly sound
related, the majority are not.

The second data source is the commonly-used AudioSet
dataset~\cite{audioset}, which consists of approximately 2 million 10-second
audio clips. Here, we consider three text annotation types:  (i) standard
human-readable labels associated with each clip in the released dataset, which
covers a fairly comprehensive set of 527 sound event categories (though the
natural language diversity here is limited with only one or two label variations
per class); (ii) a set of human-provided natural language sound
descriptions we collected for approximately 50K of the AudioSet clips; and (iii)
the AudioCaps dataset~\cite{bib:audiocaps}, which includes natural language
descriptions for approximately 46K AudioSet clips.

The final, much smaller source is a collection of
approximately 110K sound event/scene recordings from the Pro Sound Effects Library~\cite{bib:prosound}.  Each recording is paired with
a textual description, which can be anything from an unstructured list of tags
to a multiple-sentence natural language description.  


For the \tuneenv~model we also used a combination of datasets. The Clotho dataset \cite{drossos2020clotho} containing $5929$ audio clips each with 5 different captions describing them is ideally suited for this task. It is split into development (containing $3840$ clips), evaluation (containing $1045$ clips) and validation (containing $1045$ clips) subsets. The next dataset we included is the AudioCaps dataset~\cite{bib:audiocaps}, containing 38118 clips for training and 979 clips for evaluation.

Additionally we used recordings of various types of sounds provided by users via Freesound. We used the audio from the FUSS dataset~\cite{wisdom2021fuss}, which uses clips from FSD50k~\cite{fonseca2022fsd50k}. FSD50k provides manually-verified labels for 200 classes, a subset of the AudioSet ontology~\cite{audioset}.
FUSS uses a subset of FSD50k clips that are CC0-licensed and annotated with only a single label, and we also use these labels as the audio description.

For the background audio, which requires neither labels nor captions (see Section~\ref{sec:train-tune-env}), we used the FSD50k dataset~\cite{fonseca2022fsd50k} and during training also 1 million 10-second audio clips from internet videos.

\begin{figure*}[!ht]
     \centering
        \includegraphics[width=0.98\textwidth]{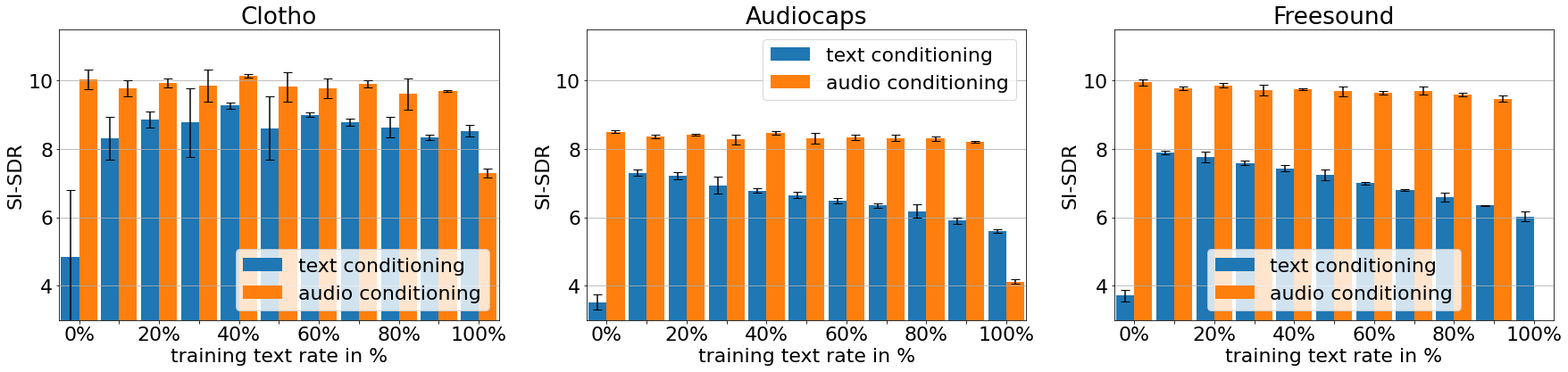}
          \vspace{-8pt}
        \caption{Enhancement SI-SDR mean and standard deviation (error bars) experimental results of using different proportions of text conditioning vs audio conditioning during training measured on the Clotho, Audiocaps and Freesound evaluation sets.}
             \vspace{-4pt}
        \label{fig:resuls}
\end{figure*}

%




\begin{table*}[!ht]
\centering
\begin{tabular}{lcc|cc|cc|cc}
  &   \multicolumn{2}{c|}{Clotho enhanced}  &   
  \multicolumn{2}{c|}{Clotho suppressed} &  \multicolumn{2}{c|}{Audiocaps enhanced} &
    \multicolumn{2}{c}{Audiocaps suppressed}\\ 
 & \textbf{audio} & \textbf{text}   & \textbf{audio} & \textbf{text}    & \textbf{audio} & \textbf{text} & \textbf{audio} & \textbf{text}\\ \hline
 \tuneenv &  $10.1 \pm 0.1$  &  $9.1 \pm 0.1$ &
 $10.0  \pm 0.1$ &  $9.0 \pm 0.1$  &
 $8.2  \pm 0.1$ &  $6.5 \pm 0.1$ &
 $8.1  \pm 0.1$ &  $6.3 \pm 0.1$ \\ 
 
 - consistency loss & $9.5 \pm 0.2$   &   $8.3 \pm 0.6 $   &
 $9.4  \pm 0.2 $ &  $8.3 \pm 0.6$ &
 $7.9  \pm 0.2$ &  $6.8 \pm 0.2$ &
 $7.8  \pm 0.2$ &  $6.6 \pm 0.2$\\

 - suppressed audio output & $9.0 \pm 0.1$   &   $6.9 \pm 0.2$   &
 - &  -  &
 $7.8  \pm 0.2$ &  $6.6 \pm 0.1$  &
 - &  -  \\
 No shared embedding \cite{gfeller2021one} &  $9.9 \pm 0.1 $ &   -  & 
 $9.9  \pm 0.1$ &  - &
 $8.4  \pm 0.1$ &  -  &
  $8.4  \pm 0.1$ &  -  
\end{tabular}
\caption{Mean SI-SDR results of our ablation experiments for \tuneenv, evaluated on the Clotho and Audiocaps datasets, using a training text rate of 0.3. We report the mean and standard deviation over three runs.}
\label{tab:results}
             \vspace{-8pt}

\end{table*}

\subsection{Training setup}

\subsubsection{\tuneenv}
\label{sec:train-tune-env}
To train the \tuneenv\ model we use audio data from two data sources: unlabeled \textit{background audio} clips and  \textit{labelled or captioned audio} clips. In this context background audio refers to undesired audio and could contain
any sounds present in the training datasets.
Each training example is constructed from one clip of each data source. For the target audio we take a 2s random crop from captioned audio. This is added to a 2s random crop from the background audio with a random gain between -4dB and +4 dB to create the mixed audio input to the model.
A second non-overlapping 2s random crop is taken from the captioned audio and is used as the conditioning audio.

During training, we randomly choose to either use text or audio as conditioning.
The balance between these two types of conditioning embeddings is controlled by the \textit{training text rate} parameter. 
In either case, the embedding is computed using the pre-trained \soundwords\ model, using either the text or the audio stack.

The \tuneenv\ model is trained for 1 million steps on 32 Google Cloud TPU v3 cores using a batch size of 512. Given pairs of mixed audio clips and conditioning vectors, it outputs two audio clips of the same length as the mixed audio input. The first output is intended to contain the audio described by the conditioning, which we refer to as the ``enhanced'' or ``tuned'' audio. We refer to the second output as the ``suppressed'' audio, which should contain all background sounds not described by the conditioning vector.

We compute three losses: two reconstruction losses and one consistency loss. One reconstruction loss is between the suppressed output and the background audio clip, and the other loss between the enhanced output and the target audio clip. For this we use scale-invariant signal-to-distortion ratio (SI-SDR)~\cite{leroux2018sdr} which we convert into a loss measure as follows:
\begin{equation}
\text{SI-SDR loss}(x, y) = -\text{softplus}(30 - \text{SI-SDR}(x, y))
\end{equation}
The consistency loss is computed between the mixed audio input clip and the sum of both output clips, and induces the model to decide which parts of the audio belong to which output. For this we use the standard signal-to-noise ratio that we convert into a loss in the analogous way to SI-SDR. Checkpoints are written every 50k steps and after training the best checkpoint that minimizes the vggish \cite{vggish} embedding distance is selected.

\subsubsection{\soundwords}

We warm-start \soundwords\ training from separate audio and text encoder models.
For the audio tower, we use a pretrained Resnetish AudioSet
classifier~\cite{bib:shershey}, where the final classifier layer is replaced
with a fully connected layer that maps to the 64-dimensional shared embedding
space.  For the text tower, we initialize with the publicly available uncased
BERT-Base embedding model~\cite{bib:bert}, again terminating with a final
fully-connected layer with output dimension 64.  After initialization, all
parameters in both towers are updated during training.  Each batch is
constructed with a preset mix of the training data sources: 30\% from the 50M
video set, 5\% from our 50K collected captions, 25\% from AudioCaps, 10\% from
AudioSet labels, and 30\% from Pro Sound (no optimization of these proportions
was performed).  The loss is computed over all examples across multiple
accelerators, leading to an effective batch size of 6144 target audio-text
pairs.  We use the Adam optimizer with learning rate 5e-5 and a trainable
softmax temperature with initial value 0.1 (details in~\cite{bib:cmc}).

\section{Evaluation}
\label{sec:eval}
In this section we first evaluate the effectiveness of the {\soundwords} model for its intrinsic purpose of joint embedding, followed by a demonstration of its usefulness in providing conditioning signals for {\tuneenv}. As the default separation evaluation metric, we use SI-SDR, as is commonly done in traditional audio separation tasks. When evaluating using audio conditioning we extract the target audio clip and the conditioning audio clip from non-overlapping crops of the same input example. In all cases we perform three training runs and report the mean and standard deviation across these runs.
Audio examples produced by our model are provided online\footnote{\url{https://google-research.github.io/seanet/textsoundfilter/examples/}}.

\subsection{SoundWords}
In order to substantiate that our weakly-supervised joint embedding model
performs reasonably, we borrowed the Clotho-based text-to-audio retrieval
evaluation defined in the prior attempt at training such a
model~\cite{bib:vgg}. The Clotho dataset was originally designed for audio captioning
research, but here it is repurposed.  We embed each of the 5225 captions and 1045
audio clips included in the Clotho evaluation set (each clip is associated with
5 unique captions).  For each caption, we retrieve the closest audio clips in the
shared embedding space and measure ``recall at 1'' (R@1), the fraction of captions
whose associated clips are in the top position.  
Unlike the past study~\cite{bib:vgg}, we perform this task in a zero-shot
capacity: while we train on a large-scale collection of audio-text pair data
sources, we do not use the Clotho training dataset in any capacity.  Instead we
evaluate our model directly on the Clotho evaluation set.  Despite the
disadvantage of a train-test domain shift, we still reach an R@1 retrieval
performance of 5.3\%, which lies in the 4.0\%--9.6\% range reported in the
prior study (all of which relied on within-domain training on Clotho and
various pretraining strategies).

\subsection{SoundFilter}

The proposed model achieves a mean enhancement SI-SDR of 10.1 (9.1) and a mean suppressed SI-SDR of 10.0 (9.0) on the Clotho dataset when evaluated using audio (text) conditioning. In a set of ablation experiments, we demonstrate the effects of the individual components of our method, see Table~\ref{tab:results}. We see a high correlation between a SI-SDR of a model's enhancement output and its suppression output.




The most import parameter in our training setup is \textit{training text rate}, which controls the percentage of training examples that condition with text instead of audio. For 0\%, the \tuneenv~model is only trained using audio conditioning, and for 100\%, the model is only trained with text conditioning. In the first experiment we vary this parameter.  Figure~\ref{fig:resuls} shows the SI-SDR in dB measured between the target audio clip and the enhanced output on the Clotho, Audiocaps and Freesound evaluation datasets using either only text conditioning or only audio conditioning at test time. Unsurprisingly, when we train only on audio conditioning (0\%  training text rate), the model performs worst on text conditioning evaluations, and when trained only on text conditioning (100\%  training text rate), the model performs worst on audio conditioning evaluations. Between these two extremes we can observe three things:
\begin{enumerate}
    \item The best models for text conditioning are not those with a training text rate 100\%, but are those with a training text rate of around 10\% - 40\%.
    \item Varying the training text rate between 0\% and 90\% does not seem to affect the performance of the audio conditioned evaluations much.
    \item Results obtained using audio conditioning at test time typically achieve a higher level of SI-SDR  than results obtained using text conditioning.
\end{enumerate}

These observations might be explained by the data that we used. We only have a limited number of audio clips with useful labels or captions, and while different random crops and gains allow us to create many different audio conditioning examples, we can only obtain a single text conditioning example. This can lead to overfitting, especially when the \textit{training text rate} parameter is high.

By evaluating using both audio conditioning and target clips from the same audio clip,
we suspect the model does not need to generalize as much compared to using text conditioning. For example, text conditioning may be something general such as \textit{engine starting up}, while audio conditioning for the same example would likely contain the same engine as the target audio. We suspect this is why overfitting does not affect the audio conditioned evaluations as much and explains why they generally achieve higher SI-SDR than text conditioning. Additionally, the textual descriptions do not always exhaustively describe the associated audio and may omit some sound sources that would be present in the audio conditioning extracted from the same clip.



\section{Conclusions}
In this paper, we proposed the first text-driven universal sound separation model. Our model combines \soundwords, a contrastively-trained text-audio embedding model, and \tuneenv, a conditional sound separation model. We demonstrated the effectiveness of our model via extensive evaluation on three separation datasets constructed from audio annotated with textual descriptions or class labels. Future work will investigate the effectiveness expanding our conditioning modalities to include images.


\bibliographystyle{IEEEtran}

\bibliography{references}

\end{document}